\documentclass[runningheads, envcountsame, a4paper]{llncs}

\usepackage{amssymb,amsmath}
\usepackage{mathtools}
\usepackage{cite}
\usepackage{tikz}
\usepackage{enumerate}

\usepackage{tikz}
\newcounter{braid}
\newcounter{strands}
\pgfkeyssetvalue{/tikz/braid height}{0.5cm}
\pgfkeyssetvalue{/tikz/braid width}{0.5cm}
\pgfkeyssetvalue{/tikz/braid start}{(0,0)}
\pgfkeyssetvalue{/tikz/braid colour}{black}
\pgfkeys{/tikz/strands/.code={\setcounter{strands}{#1}}}

\makeatletter
\def\cross{%
  \@ifnextchar^{\message{Got sup}\cross@sup}{\cross@sub}}

\def\cross@sup^#1_#2{\render@cross{#2}{#1}}

\def\cross@sub_#1{\@ifnextchar^{\cross@@sub{#1}}{\render@cross{#1}{1}}}

\def\cross@@sub#1^#2{\render@cross{#1}{#2}}

\def\render@cross#1#2{
  \def\strand{#1}
  \def\crossing{#2}
  \pgfmathsetmacro{\cross@y}{-\value{braid}*\braid@h}
  \pgfmathtruncatemacro{\nextstrand}{#1+1}
  \foreach \thread in {1,...,\value{strands}}
  {
    \pgfmathsetmacro{\strand@x}{\thread * \braid@w}
    \ifnum\thread=\strand
    \pgfmathsetmacro{\over@x}{\strand * \braid@w + .5*(1 - \crossing) * \braid@w}
    \pgfmathsetmacro{\under@x}{\strand * \braid@w + .5*(1 + \crossing) * \braid@w}
    \draw[braid] \pgfkeysvalueof{/tikz/braid start} +(\under@x pt,\cross@y pt) to[out=-90,in=90] +(\over@x pt,\cross@y pt -\braid@h);
    \draw[braid] \pgfkeysvalueof{/tikz/braid start} +(\over@x pt,\cross@y pt) to[out=-90,in=90] +(\under@x pt,\cross@y pt -\braid@h);
    \else
    \ifnum\thread=\nextstrand
    \else
     \draw[braid] \pgfkeysvalueof{/tikz/braid start} ++(\strand@x pt,\cross@y pt) -- ++(0,-\braid@h);
    \fi
   \fi
  }
  \stepcounter{braid}
}

\tikzset{braid/.style={double=\pgfkeysvalueof{/tikz/braid colour},double distance=1pt,line width=2pt,white}}

\newcommand{\braid}[2][]{%
  \begingroup
  \pgfkeys{/tikz/strands=2}
  \tikzset{#1}
  \pgfkeysgetvalue{/tikz/braid width}{\braid@w}
  \pgfkeysgetvalue{/tikz/braid height}{\braid@h}
  \setcounter{braid}{0}
  \let\sigma=\cross
  #2
  \endgroup
}
\makeatother

\newcommand{\Z}{\mathbb{Z}}
\newcommand{\A}{\mathcal{A}}
\newcommand{\B}{\mathcal{B}}
\newcommand{\PCP}{\operatorname{PCP}}
\newcommand{\wPCP}{\omega\operatorname{PCP}}
\newcommand{\seq}[1]{\left<#1\right>}
\newcommand{\bx}{\mathbf x}
\newcommand{\bu}{\mathbf u}
\newcommand{\bv}{\mathbf v}

\title{Weighted Automata on Infinite Words in the Context of Attacker-Defender Games}
\titlerunning{Weighted Automata on Infinite Words and Attacker-Defender Games}
\toctitle{Weighted Automata on Infinite Words in the Context of Attacker-Defender Games}

\author{Vesa Halava\inst{1}\fnmsep\inst{2} \and Tero Harju\inst{1} \and Reino Niskanen\inst{2}\fnmsep\thanks{The author was partially supported by Nokia Foundation Grant} \and Igor Potapov\inst{2}\fnmsep\thanks{
The author was partially supported by EPSRC grant ``Reachability problems for words, matrices and maps'' (EP/M00077X/1)
}}

\authorrunning{V.~Halava, T.~Harju, R.~Niskanen, and I.~Potapov}
\tocauthor{Vesa~Halava, Tero~Harju, Reino~Niskanen, and Igor~Potapov}

\institute{Department of Mathematics and Statistics \\ 
University of Turku,
FIN-20014 Turku, Finland \\
\email{$\{$vesa.halava,harju$\}$@utu.fi}
\and
Department of Computer Science, 
University of Liverpool \\  
Ashton Building, Liverpool, L69 3BX, UK \\
 \email{$\{$r.niskanen,potapov$\}$@liverpool.ac.uk}
	}

\begin{document}
\maketitle
\setcounter{footnote}{0}

\begin{abstract}
We consider several infinite-state Attacker-Defender games with reachability objectives. 
The results of the paper are twofold. Firstly we prove a new language-theoretic result for 
weighted automata on infinite words and show its encoding into the framework of Attacker-Defender games.
Secondly we use this novel concept to prove undecidability for checking existence of 
a winning strategy in several low-dimensional mathematical games including 
vector reachability games, word games  and braid games.
 \end{abstract}

\section{Introduction}
In the last decade there has been a steady, growing interest in the area of infinite-state games and computational complexity of the 
problem of checking the existence of a winning strategy 
\cite{ArulReichert,CSL2003,BJK2010,KF-CSL2013,RobotGames,KV2000,W2001}.
Such games provide 
powerful mathematical framework for a large number of computational problems. In particular they appear 
in the verification, refinement, and compatibility checking of reactive systems \cite{AHK-ACM02},
analysis of programs with recursion \cite{KF-CSL2013}, combinatorial topology and have deep connections with automata theory and logic 
\cite{R69,KV2000,W2001}.
In many cases the most challenging problems appear in low-dimensional models or systems, where it is likely to have 
a few special cases with decidable problems and open general problem 
as the system may produce either too complex behaviour for analysis or a lack of 
``space'' to code directly the universal computation for showing undecidability of the problem.

In this paper we present three variants of low-dimensional Attacker-Defender games
(i.e. 
Word Games, Matrix Games and Braid Games)
for which it is undecidable to determine whether one of the players has a winning strategy.  
In addition the proof incorporates new language theoretical result (Theorem \ref{UniversalFA}) about weighted automata on infinite words that can be efficiently used in the context of other reachability games.

The Attacker-Defender game is played in rounds, where in each round  the move of Defender (Player 1) is followed by the move of Attacker (Player 2) starting from some initial position. Attacker tries to reach a target position while Defender tries to keep Attacker from reaching the target position. Then we say that Attacker has a winning strategy if it can eventually reach a target position regardless of Defender's moves. We show that in a number of restricted cases of such games it is not possible to decide about 
existence of the winning strategy for a given set of moves, initial and target positions.  

We show that if both players are stateless but the moves correspond to a very restricted linear transformation from $SL(4,
\Z)$ the problem of existence of  winning strategy is undecidable. One can show that using a direct translation from known undecidable reachability games (Robot Games \cite{RobotGames}) leads to undecidability for linear transformations in dimension 18.
To prove it we first generalize the concept by introducing the {\sl Word Game},
where players are given words over a group alphabet and in alternative way concatenate their words with a goal 
for Attacker to reach the empty word.  The games on words are common for proving results in language theory
\cite{Kunc2005,Kunc2007,LW2014} over semigroup alphabets, while we formulate a game with much simpler reachability objective 
for games over a group alphabet. 

Later we show that it is possible to stretch the application of the proposed techniques to other 
models and frameworks even looking at the games on topological objects, which were recently studied in \cite{CDW, BC-preprint}.
Braids are classical topological objects that attracted a lot of attention
due to their connections to topological knots and links 
as well as their applications to polymer chemistry, molecular biology,
cryptography, quantum computations and robotics \cite{GC2006,Dehornoy, EPCHLT92,G,PP2011}.
In this paper we consider games on braids with only 3 or 5 strands, where the braid is modified by composition of 
braids from a finite set with a target for Attacker to reach a trivial braid.  We show that it is undecidable to check the existence of a winning strategy for 3 strands from a given nontrivial braid and for 5 strands starting from a trivial braid, while the reachability with a single player (i.e. with nondeterministic concatenation from a single set) was shown to be decidable for braids with 3 strands in \cite{P-FSTTCS13}.

The whole paper is also based on another important language-theoretic result showing that 
the universality problem for weighted automata $\A$ having merely five states accepting infinite words is undecidable.
The acceptance of an infinite word $w$ means that there exists a finite prefix $p$ of $w$ 
such that for a word $p$  there is a path in $\A$ that has the zero weight. 
The problem whether all infinite words are accepted for a given $\A$ is undecidable and corresponds to the fact that 
there is no solution for infinite PCP.  

The considered model of automaton is closely related to the \emph{integer
weighted finite automata}  as defined in~\cite{halavaharju1} and \cite{ABK2011}, where finite
automata are accepting finite words and having additive integer weights on the
transitions. In~\cite{halavaharju1} it was shown that the universality
problem is undecidable for integer weighted finite automata on finite words by reduction from Post Correspondence Problem.
In the context of a game scenario it is important to have a property of acceptance in relation to infinite words
with a finite prefix reaching a target value. 
Our proof of undecidability in this paper initially follows the idea from~\cite{halavaharju1}
for mapping computations on words into weighted (one counter) automata model.

Complete proofs can be found in the Appendices.

\section{Notations and Definitions}

An \emph{infinite word} $w$ over a finite alphabet $A$ is an infinite
sequence of letters $w=a_0a_1a_2a_3\cdots$ where $a_i \in A$ is a letter
for each $i=0,1,2,\ldots$. We denote the set of all infinite
words over $A$ by $A^\omega$. The monoid of all finite words over
$A$ is denoted by $A^*$.
A word $u \in A^*$ is a \emph{prefix} of $v \in A^*$, denoted by $u \leq v$, if
$v=uw$ for some $w \in A^*$. If $u$ and $w$ are both nonempty, then the prefix $u$ is called \emph{proper}, denoted by $u < v$.
A \emph{prefix} of an infinite word $w \in A^\omega$ is a finite word $p \in A^*$
such that $w=pw'$ where $w'\in A^\omega$. This is also denoted by $p\leq w$.
The length of a finite word $w$ is denoted by $|w|$. For a word $w$, we denote by
$w(i)$ the $i$th letter of $w$, i.e.,
$w=w(1)w(2) \cdots$. Let  $w=w(1)\cdots w(n)$, its \emph{reversed word} is denoted by $w^R=w(n)\cdots w(1)$, i.e. the order of the letters is reversed.

Consider a finite integer weighted automaton $\A = (Q,A,\sigma, q_0,F,\Z)$
with the set of states $Q$, the finite alphabet $A$, the set of transitions
$\sigma\subseteq Q\times A\times Q\times \Z$, the initial state $q_0$, the set of final states $F\subseteq Q$ and the additive group of integers $\Z$ with identity $0$, that is $(\Z,+,0)$. 
We write the transitions in the form
$t=\seq{q,a,p,z}\in\sigma$. In the graphical presentation an edge $t$ is denoted by $q\stackrel{(a,z)}{ \ \longrightarrow \ }p$.
Note that $\A$ is non-deterministic complete automaton in a sense that for each $q\in Q$ and $a\in A$ there is atleast one transition $\seq{q,a,p,z}\in\sigma$ for some $p\in Q$ and $z\in\Z$. 

Let $\pi = t_{i_0} t_{i_1} \cdots$ be an infinite
path of $\A$, where $t_{i_j} = \seq{q_{i_j},a_j,q_{i_{j+1}},z_j}$ for
$j\ge 0$. 
Define the morphism $\|\cdot \| \colon \sigma^\omega \to A^\omega$ by setting $\| t \| =
a$ if $t=\seq{q,a,p,z}$.
Let $p=t_{i_0}t_{i_1}\cdots t_{i_n}$ for some $n$ be a prefix of $\pi$.
The \emph{weight of the prefix} $p$ is 
$
\gamma(p) = z_0+z_1+\ldots+z_n\in \Z .
$
The prefix $p$ \emph{reaches} state $q\in Q$  if the last transition
of $p$ enters~$q$, i.e., if  $t_n=(q_{i_n},a_n,q_{i_{n+1}},z_n)$, then
$q_{i_{n+1}}=q$. Denote by $R(p)$ the state reached by the finite path $p$.


An infinite word $w\in A^\omega$ is accepted by $\A$ if there exists an infinite path $\pi$
such that at least one prefix $p$ of $\pi$ reaches a state in $R(p) \in F$ and 
has weight
$\gamma(p)=0$. The language \emph{accepted by} $\A$ is
\begin{align*}
L(\A) &= \left\{w\in A^\omega \mid \exists \pi\in \sigma^\omega\colon ||\pi||=w \right. \text{ and } \\
& \qquad \left. {} \exists \text{ prefix } p \text{ of } \pi\colon \gamma(p)=0 \text{ and } R(p)\in F\right\}.
\end{align*}
We also define \emph{reverse acceptance}, used in undecidability in Attacker-Defender games, in which instead of prefix $p$ we consider $p^R$ and its weight.
Now an infinite word $w$ is accepted by the automaton if and only if for corresponding computation $\pi$ there exists a prefix whose reverse has zero weight. That is,
\begin{align*}
L_R(\A) &= \left\{w\in A^\omega \mid \exists \pi\in \sigma^\omega\colon ||\pi||=w \right. \text{ and } \\
& \qquad \left. {} \exists \text{ prefix } p \text{ of } \pi\colon \gamma(p^R)=0 \text{ (and } R(p)\in F)\right\}.
\end{align*}

A \emph{configuration} of $\A$ is any triple $(q,w,z) \in Q\times A^* \times \Z$.
A configuration $(q,aw,z_1)$ is said to \emph{yield} a configuration
$(p,w,z_1+z_2)$, denoted by
$
(q,aw,z_1) \models_{\A} (p,w, z_1+z_2),
$
if there is a transition $t=\seq{q,a,p,z_2}\in\sigma$. Let
$\models_{\A}^*$ or simply $\models^*$, if $\A$ is clear from the context, be
the reflexive and transitive closure of the relation $\models_{\A}$.
\footnote{While we restrict ourselves to the case, where the weights of the automaton are elements of the additive group of integers $\Z$, we could define the model for any other group $(G,\cdot,\iota)$ as well.}

The \emph{Universality Problem} is a problem to decide whether the language accepted by weighted automaton $\A$ is the set of all infinite words. In other words, whether or not $L(\A)=A^\omega$. The problem of \emph{non-universality} is the complement of universality problem, that is, whether or not $L(\A)\neq A^\omega$ or whether there exists $w\in A^\omega$ such that for every path $\pi$ corresponding to computation of $w$ and every prefix $p$ of $\pi$, $\gamma(p)\neq0$.

An \emph{instance} of the \emph{Post Correspondence Problem} ($\PCP$, for short)
consists of two morphisms $g,h: A^*\rightarrow B^*$, where $A$ and $B$ are
alphabets. A nonempty word $w\in A^*$ is a solution of an instance $(g,h)$ if it satisfies $g(w)=h(w)$.
It is undecidable whether or not an instance of the $\PCP$ has a solution;
see~\cite{P46}. Also the problem is undecidable for domain alphabets~$A$ with $|A|\geq 7$;
see~\cite{MS05}.
The cardinality of the domain alphabet $A$
is said to be the \emph{size} of the instance.

The \emph{Infinite Post Correspondence Problem}, $\wPCP$, is a natural extension of the $\PCP$.
An infinite word $w$ is a \emph{solution} of the instance
$(g,h)$ of the $\wPCP$ if for every finite prefix $p$ of $w$ either
$h(p)< g(p)$ or $g(p)< h(p)$. In the $\wPCP$ it is asked whether or not a given
instance has an infinite solution or not. Note that in our formulation prefixes have to be proper. It was proven in \cite{HH06} that the problem is undecidable for domain alphabets $A$ with $|A|\geq9$ and in \cite{DL12} it was improved to $|A|\geq8$. In both proofs more general formulation of $\wPCP$ was used, namely the prefixes did not have to be proper. It is easy to see that adding a new letter $\alpha$ to the alphabets and desynchronizing the morphisms $h,g$, gives us solution where prefix has to be proper. That is, we add $\alpha$ to the left of each letter in the image under $h$, to the right of each letter in the image under $g$ and $g(\alpha)=\alpha, h(\alpha)=\varepsilon$. Now the solution has to start with $\alpha$ and images cannot be of equal length because the image under $g$ ends with $\alpha$ but not under $h$. Note that in fact, both constructions already have this property, see \cite{HH06,DL12}.

\section{Universality for Weighted Automata on $A^\omega$}
We prove that the universality problem is undecidable for integer weighted
automata on infinite words by reducing the instances of the \emph{infinite Post
Correspondence Problem} to the universality problem. 

Let $(g,h)$ be a fixed instance of the $\wPCP$.
Then $g,h\colon A^*\to B^*$ where $A=\{a_1,a_2,\ldots,a_{m-1}\}$ and $B=\{b_1,b_2,\dots, b_{s-1}\}$. We construct
an integer weighted automaton $\A=(Q,A,\sigma,q_0,\{q_4\},\Z)$, where $Q=\{q_0,q_1,q_2,q_3,q_4\}$,
corresponding to the instance $(g,h)$ such that an infinite word $w \in A^\omega$ is accepted by $\A$ if and only if for some finite prefix $p$ of $w$, $g(p)\nless h(p)$ and $h(p)\nless g(p)$. Note that our automaton is complete, i.e there is a transition labeled with $(a,z)$ from each state $q_i$ for every $a\in A$ and some $z\in\Z$.

The idea of encoding $\wPCP$ and proof of undecidability for universality problem is based on computation in weighted automaton that can be partitioned into four parts A,B,C and D. Let us consider the case where the image under $h$ is always longer than the image under $g$, the other cases are taken into account in the construction of the automaton. In part A, differences of lengths of images under $h$ and $g$ are stored for initial part of the input word. In part B, differences of position $k$ of image of a letter under morphism $h$ and length of image under $g$ are stored together with a natural number $j_k$ representing letter at $k$th position. In part C, the lengths of images under morphism $g$ catch-up (by subtracting lengths of images under morphism $g$) to position specified after parts A and B. Finally, in part D, position $\ell$ in the image of the second morphism is subtracted together with a natural number $i_\ell$ from a set of natural numbers representing letters not at $\ell$th position under morphism $g$.

\begin{figure}
\begin{center}
\input{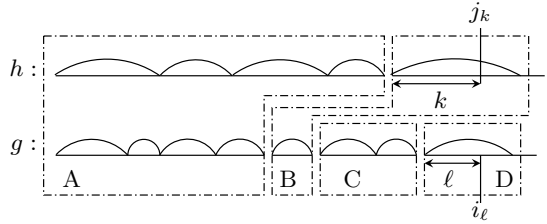}
\end{center}
\caption{Illustration of computation of $\wPCP$}
\end{figure}

To store two different values (differences of lengths and a code for a stored symbol) at the same time, we can use a single counter since we only store one symbol from a finite alphabet. Let us assume that symbols are encoded as natural numbers in $\{1,\ldots,s-1\}$, where $s$ is larger than the size of image alphabet $B$. The length $n$ will be defined as $n\cdot s$ and we have enough space to store a single symbol from $B$ by adding its code. If in the part D we refer to the same position the difference of the lengths of images should be 0 and if the letters are the same, the difference of letter codes is non-zero. This is done by allowing to subtract only that number which does not equal the number corresponding to a code of the letter at the $\ell$th position.

In the above consideration we considered the case where images under $h$ was always longer than images under $g$. To make the construction work for all cases several computation paths are needed to be implemented in the automaton. The difference in lengths of images is positive when image under $h$ is longer and negative when image under $g$ is longer. For each case there are two possibilities for position of error. Either the difference is small enough that, after reading the next letter, there will be a position in images where letters differ (parts B and D have to be done simultaneously), or the difference is large enough, that image of the second morphism has to catch-up before error can be verified. Also from our formulation of $\wPCP$, it is possible that images are of equal length which means that the word is not a solution of $\wPCP$.

\begin{figure}
\begin{center}
\scalebox{0.8}{\input{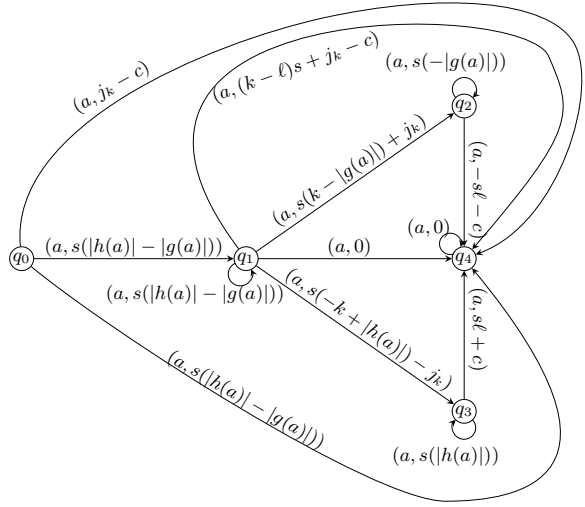}}
\end{center}
\caption{\label{fig:FA}The weighted automaton $\A$.
In the figure  $a\in A$.}
\end{figure}

\begin{lemma}\label{SolLang}
Let $w\in A^\omega$. Then $w$ is a solution of an instance $(g,h)$ of the $\wPCP$ if and only if $w\notin L(\A)$.
\end{lemma}

\begin{theorem}\label{UniversalFA}
It is undecidable whether or not $L(\A)=A^\omega$ holds for integer weighted automata $\A$ over its alphabet $A$. 
\end{theorem}
\begin{proof}
Claim follows from Lemma~\ref{SolLang} and the undecidability of infinite $\PCP$, see \cite{R85}. The full construction of automaton $\A$ can be found in Appendix A and is depicted in Figure~\ref{fig:FA}. 
\qed
\end{proof}

\begin{corollary}\label{NonUniversalFA}
It is undecidable whether or not for weighted automaton $\A$, there exists a word $w\in A^\omega$ such that for its each computation path $\pi$ and prefix $p\leq\pi$, $\gamma(p)\neq0$ holds.
\end{corollary}

For proving undecidability of finding a winning strategy in Attacker-Defender games we need to utilize slightly different acceptance condition.

\begin{theorem}\label{UndecidableReverseAcceptance}
It is undecidable whether or not $L_R(\B)=A^\omega$ holds for integer weighted automata $\B$ over its alphabet $A$.
\end{theorem}
\begin{proof}[Sketch]
The proof is based on Theorem~\ref{UniversalFA}. It is easy to see that if we reverse all edges and follow the computation path $p$ of original automaton from end to finish, we have a computation path for $p^R$. The full proof is in Appendix A.
\end{proof}

\section{Applications to Attacker-Defender Games}
Let us consider a two-player Attacker-Defender game which is played in rounds
and in each round a move of Defender is followed by a move of Attacker starting from some initial position.   Attacker tries to reach a target position while Defender tries to keep Attacker from reaching the target position. Attacker has a winning strategy if it can eventually reach a target position regardless of Defender's moves.
The main computational question is to check whether Attacker has a winning strategy for a given set of moves, initial and target positions.  

Following the result for the weighted automata on infinite words  (Theorem~\ref{UndecidableReverseAcceptance}) we can now define 
a simple scenario of undecidable infinite-state game that can be also applied to other game frameworks.
Assume that Defender will provide any input letters from a finite alphabet, one by one, to Attacker and Attacker appends dummy symbol $\#$ until he chooses to follow computation path of automaton $\B$ of Theorem~\ref{UndecidableReverseAcceptance}. Attacker has to decide whether provided word (ignoring symbols $\#$) when played according to available transitions is accepted by $\B$. 

In the above framework Defender will have a winning strategy if there is a 
solution for infinite PCP and Attacker will have a winning strategy otherwise.

\subsection{Weighted Word Game}
%
Let us define the Attacker-Defender game on words, where the moves of Attacker and Defender correspond to
concatenations of words (over free group alphabet) and follows a computation path of weighted automaton. This simplification allows us to use {\sl Word Game} to prove nontrivial results for 
games with {\sl low-dimensional} linear transformations and topological objects just by using  injective homomorphism (i.e. monomorphism) to map words into other mathematical objects.

A \emph{weighted Word Game}  consists of two players, \emph{Attacker} and \emph{Defender} having sets of words  $\{u_1,\ldots,u_r\}\subseteq \Gamma^*$ and $\{v_1,\ldots,v_s\}\subseteq\Gamma^*$ respectively, where $\Gamma$ is finite alphabet from a free group, and integers $x_{u_1},\ldots,x_{u_r},x_{v_1},\ldots,x_{v_s}$ corresponding to each word.
In each round Defender chooses the word before Attacker, the initial position is the pair $(w,0)$, where $w\in\Gamma^*$ and 0 is initial value of the counter, and target position of this game is the group identity, i.e. the empty word, with zero weight.
The \emph{configuration} of a game at time $t$ is denoted by a word  $w_t$ and integer $x$ as a counter. In each round of the game both Defender and Attacker concatenate their words (append from the right) and update the counter value.  
Clearly $w_t=w\cdot v_{i_1} \cdot u_{i_1} \cdot v_{i_2} \cdot u_{i_2} \cdot \ldots \cdot v_{i_t} \cdot u_{i_t}$ 
after $t$ rounds of the game, where $u_{i_j}$ and $v_{i_j}$ are words from defined above sets of Attacker and Defender, and the counter value is $\sum_{j=1}^t (x_{v_{i_j}}+x_{u_{i_j}})$.
The decision problem for the word game is to check whether there exists a winning
strategy for Attacker to reach an empty word with zero weight. 

\begin{lemma}\label{groupEnc} Let $\Sigma' = \{z_1, z_2, \ldots, z_l\}$ be a group alphabet and $\Sigma_2=\{c,d,\overline{c},\overline{d}\}$ be a binary 
group alphabet. Define the mapping $\alpha:\Sigma' \to \Sigma_2^*$ by:
$\alpha (z_i) = c^i d\overline{c}^{i}, \alpha (\overline{z_i}) = c^i\overline{d}\overline{c}^{i}, $
where $1 \leq i \leq l$. Then $\alpha$ is a monomorphism. Note that $\alpha$ can be extended to
domain $\Sigma'^*$ in the usual way \cite{BM07,Paul_Igor_FI_2012}.
\end{lemma}

\begin{theorem}\label{weighted Word Game}
It is undecidable whether Attacker has a winning strategy in the weighted Word Game with words over a binary free group alphabet.
\end{theorem}
The idea of proof is that Defender plays words of $\{v_1,\ldots,v_s\}$ corresponding to $w\in A^\omega$ letter by letter. Attacker either plays $\#$ or starts following computation path of the automaton $\B$ and stores weights in a counter. Word $q_0\overline{q_4}$, where $q_0$ and $q_4$ are initial and final states of $\B$ respectively, with weight 0 is reached if and only if $w$ is accepted by the automaton. Then we encode the group alphabet using Lemma~\ref{groupEnc} to have binary group alphabet. The full proof can be found in \cite{HHNP15}.

\subsection{Word Games on Pairs of Group Words}
%
We now modify the game of previous section by encoding counter as a separate word over unary group alphabet $\Gamma'=\{\rho,\overline{\rho}\}$.

Now \emph{Word Game}  consists of Attacker and Defender having sets of words  $\{(u_1,u'_1),\ldots,(u_r,u'_r)\}\subseteq \Gamma^*\times\Gamma'^*$ and $\{(v_1,\varepsilon),\ldots,(v_s,\varepsilon)\}\subseteq\Gamma^*\times\Gamma'^*$ respectively, where $\Gamma$ is binary group alphabet.
Now the configuration of a game after $t$ rounds is a word  $w_t=(v_{i_1},\varepsilon) \cdot (u_{i_1},u'_{i_1}) \cdot (v_{i_2},\varepsilon) \cdot (u_{i_2},u'_{i_2}) \cdot \ldots \cdot (v_{i_t},\varepsilon) \cdot (u_{i_t},u'_{i_t})$, where $(u_{i_j},u'_{i_j})$ and $(v_{i_j},\varepsilon)$ are words from defined above sets of Attacker and Defender.  The initial position is the word $(w,\varepsilon)$ and target position of this game is an empty word $(\varepsilon,\varepsilon)$.
The decision problem for the word game is to check whether there exists a winning
strategy for Attacker to reach an empty word $(\varepsilon,\varepsilon)$.

\begin{theorem}\label{Word Game}
It is undecidable whether Attacker has a winning strategy in the Word Game with one component of words over a binary free group alphabet and the other over unary group alphabet.
\end{theorem}
\begin{proof}[Sketch] The proof is based on Theorem~\ref{weighted Word Game} with encoding of counter $x$ as a word $\rho^x$ over unary group alphabet $\{\rho,\overline{\rho}\}$. The full proof is in the Appendix B.
\end{proof}

We follow idea of \cite{BP10} to construct a word game where the initial word is $(\varepsilon,\varepsilon)$. For this we do 4 consequent games over 4 disjoint group alphabets. The games are constructed in such way that $(\varepsilon,\varepsilon)$ is reached if and only if $(\varepsilon,\varepsilon)$ is reached in every game. If none of the games result in $(\varepsilon,\varepsilon)$, then there are at least 4 words, from distinct group alphabets, which are non-canceled. Now if the computation is completed twice (i.e. 8 games have been played in total), the number of non-canceled elements cannot decrease.

\subsection{Matrix Games and Braid Games}

We extend the domain of the game and a set of rules to the class of linear transformations on integer lattice $\Z^4$ and 
to the domain of braids considering moves of the game as a concatenation of braids in $B_3$ (a class of braids with only three strands) and $B_5$ (a class of braids with only five strands)
\cite{P-FSTTCS13}.

A \emph{Matrix Game} consists of two players, \emph{Attacker} and \emph{Defender} having sets of
linear transformations  $MU_1,\ldots,MU_r\subseteq \Z^{n \times n }$ and $MV_1,\ldots,MV_s\subseteq\Z^{n \times n }$
respectively and an \emph{initial vector} $\bx_0 \in \Z^n$ of the game representing a starting position. 
The \emph{dimension} of the game is clearly the dimension of the integer lattice $n$
. 
Starting from $\bx_0$, players move the current point by applying available linear transformations (by matrix multiplication)
from their respective sets in turns. 
%
The decision problem of the {\sl Matrix Game} is to check whether there exist a winning
strategy for Attacker to return to the starting point (vector in $\Z^n$) of the game.

\begin{theorem}\label{matrixgame}
Given two finite sets matrices $MU_1,MU_2,\ldots,MU_r\subseteq \Z^{n \times n }$ and $MV_1,MV_2,\ldots,MV_s\subseteq\Z^{n \times n }$
for \emph{Attacker} and \emph{Defender} players respectively and initial starting vector $\bx_0 \in \Z^n$.
It is undecidable whether Attacker has a winning strategy in the Matrix Game of dimension four, i.e. when $n=4$.
\end{theorem}
\begin{proof}[Sketch]
We encode word game on pairs of words into matrices from $SL(4,\Z)=\{M\in \Z^{4\times 4}\mid det(M)=\pm1\}$. Identity matrix is reachable if and only if empty word is reachable in Word Game. The full proof is in the Appendix B. 
\end{proof}

Now we translate the Attacker-Defender games into games on topological objects  - braids in $B_n$. We consider very simple games on braids with only 3 or 5 strands (i.e. $B_3$ or $B_5$) where the braid is modified by composition with a finite set of braids.  We show that it is undecidable to check the existence of a winning strategy in such game, while the reachability with a single player (i.e. with nondeterministic concatenation from a single set) was shown to be decidable for $B_3$ and undecidable for $B_5$ in \cite{P-FSTTCS13}.

\begin{definition}\label{Braid_Artin}
The $n$-strand braid group  $B_n$ is the group given by the presentation with $n-1$
generators $\sigma_1 , \ldots , \sigma_{n-1}$ and the following relations
$\sigma_i \sigma_j = \sigma_j \sigma_i$, for $|i-j| \geq 2$
and $\sigma_i \sigma_{i+1} \sigma_i = \sigma_{i+1} \sigma_i \sigma_{i+1}$
for $1 \leq i \leq n-2$. These relations are called Artin's relations.
%
Words in the alphabet $\{ {\sigma}$ , ${\sigma}^{-1} \}$ will be referred to as braid words.

\end{definition}

The Braid Game can be defined in the following way. Given a set of  words for Attacker $\{a_1,...,a_r\} \subseteq B_n$ and Defender
$\{d_1,...,d_s\} \subseteq B_n$ will correspond to braids (or braid words in $B_n$). The game is starting with a given initial braid $c$ and each following configuration of the game is changed by Attacker or Defender by concatenating braids from their corresponding sets.
The concatenation (composition) of two braids is defined by putting one after the other making the endpoints of the first one coincide with the starting points of the second one. There is a neutral element for the composition: it is the trivial braid, also called identity braid, i.e. the class of the geometric braid where all the strings are straight. 

Finally, the goal of Attacker is to unbraid, i.e. to reach a configuration of the game that is isotopic to the trivial braid (empty word) and 
Defender tries to keep Attacker from reaching it.
Two braids are isotopic if their braid words can be translated one into each other via the relations from the Definition~\ref{Braid_Artin}
plus the relations $\sigma_i \sigma_i^{-1}  = \sigma_i^{-1} \sigma_i =1$, where $1$ is the identity (trivial braid).

\begin{theorem}\label{braidgame3}
The Braid Game is undecidable for braids from $B_3$ starting from non-trivial braid and for braids from $B_5$ starting from a trivial braid.
\end{theorem}
The idea is to encode words of weighted word game into braids of $B_3$ and weight into central element of $B_3$. While $B_5$ contains direct product of two free subgroup and can encode pair of words of word game into braids of $B_5$. The full proof is in Appendix B.

\bibliographystyle{splncs03}
\bibliography{mainarxiv}

\newpage
\section*{Appendix A}
We prove that the universality problem is undecidable for integer weighted
automata on infinite words by reducing the instances of the \emph{infinite Post
Correspondence Problem} to the universality problem. 

Let $(g,h)$ be a fixed instance of the $\wPCP$.
Then $g,h\colon A^*\to B^*$ where $A=\{a_1,a_2,\ldots,a_{m-1}\}$ and $B=\{b_1,b_2,\dots, b_{s-1}\}$. We construct
an integer weighted automaton $\A=(Q,A,\sigma,q_0,\{q_4\})$, where $Q=\{q_0,q_1,q_2,q_3,q_4\}$,
corresponding to the instance $(g,h)$ such that an infinite word $w \in A^\omega$ is accepted by $\A$ if and only if for some finite prefix $p$ of $w$, $g(p)\nless h(p)$ and $h(p)\nless g(p)$.

Let us begin with the transitions of $\A$, see Figure~\ref{fig:FA}.  (Recall that the weight function $\gamma$ is embedded in the transitions.) Recall also that  the cardinality of the alphabet $B$ is $s-1$.
First for each $a\in A$, let
\begin{eqnarray*}
\seq{q_0,a,q_1,s(|h(a)|-|g(a)|)}, \quad \seq{q_0,a,q_4,s(|h(a)|-|g(a)|)}, \\
\seq{q_1,a,q_1,s(|h(a)|-|g(a)|)}, \quad \seq{q_2,a,q_2,s(-|g(a)|)}, \\
\seq{q_3,a,q_3,s(|h(a)|)}, \quad \seq{q_1,a,q_4,0}, \quad \seq{q_4,a,q_4,0}
\end{eqnarray*}
be in $\sigma$. 
For error checking we need the following transitions for all letters $a\in A$:
Let $h(a)=b_{j_1}b_{j_2}\cdots b_{j_{n_1}}$, where $b_{j_k}\in B$, for each index
$1\le k \le n_1$. Then let, for each $k=1,\dots, n_1$, i.e. $j_k\in\{1,\ldots,s-1\}$ for all $k=1,\ldots,n_1$,
\begin{equation}\label{errdeth}
\seq{q_1,a,q_2,s(k-|g(a)|)+j_k} \in \sigma.
\end{equation}
Let $g(a)=b_{i_1}b_{i_2}\cdots b_{i_{n_2}}$, where $b_{i_\ell}\in B$, for each index
$1\le \ell \le n_2$, i.e. $j_\ell\in\{1,\ldots,s-1\}$ for all $\ell=1,\ldots,n_2$. For each $\ell=1,\dots, n_2$ and letter $b_c\in B$ such that $b_{i_\ell}\neq b_c\in B$, let
\begin{equation}\label{errverh}
\seq{q_2,a,q_4,-s\ell-c} \in \sigma.
\end{equation}
Symmetrically we define edges for $g(a)=b_{j_1}b_{j_2}\cdots b_{j_{n_3}}$, where $b_{j_k}\in B$, for each index
$1\le k \le n_3$. Then let, for each $k=1,\dots, n_3$,
\begin{equation}\label{errdetg}
\seq{q_1,a,q_3,s(-k+|h(a)|)-j_k} \in \sigma.
\end{equation}
Let $h(a)=b_{i_1}b_{i_2}\cdots b_{i_{n_4}}$, where $b_{i_\ell}\in B$, for each index
$1\le \ell \le n_4$. For each $\ell=1,\dots, n_4$ and letter $b_c\in B$ such that $b_{i_\ell}\neq b_c\in B$, let
\begin{equation}\label{errverg}
\seq{q_3,a,q_4,s\ell+c} \in \sigma.
\end{equation}

Let $h(a)=b_{j_1}b_{j_2}\cdots b_{j_{n_1}}$, where $b_{j_k}\in B$, for each index
$1\le k \le n_1$ and $g(a)=b_{i_1}b_{i_2}\cdots b_{i_{n_2}}$, where $b_{i_\ell}\in B$, for each index
$1\le \ell \le n_2$. For each $k=1,\dots, n_1$ and $\ell=1,\dots, n_2$ and letter $b_c\in B$ such that $b_{i_\ell}\neq b_c\in B$,
\begin{equation}\label{guessver1}
\seq{q_1,a,q_4,(k-\ell)s+j_k-c}\in\sigma.
\end{equation}

Finally, let  $h(a)=b_{j_1}b_{j_2}\cdots b_{j_{n_1}}$, where $b_{j_k}\in B$, for each index
$1\le k \le n_1$ and $g(a)=b_{i_1}b_{i_2}\cdots b_{i_{n_2}}$, where $b_{i_\ell}\in B$, for each index
$1\le \ell \le n_2$. For each $k=1,\ldots, \min\{n_1,n_2\}$ and letter $b_c\in B$ such that $b_{i_k}\neq b_c\in B$,
\begin{equation}\label{guessver2}
\seq{q_0,a,q_4,j_k-c}\in\sigma.
\end{equation}

We call the transitions in \eqref{errdeth} and \eqref{errdetg} \emph{error guessing transitions} and in \eqref{errverh} and \eqref{errverg} \emph{error verifying transitions}. Note that transitions in \eqref{guessver1} and \eqref{guessver2} are both error guessing and verifying transitions.

The idea is to keep track of differences in lengths of images under $g$ and $h$ multiplied by $s$ and then guess and verify an error in the images by storing letters of the image alphabet. The difference in lengths of images is positive when image under $h$ is longer and negative when image under $g$ is longer. For each case there are two possibilities for position of error. Either the difference is small enough that, after reading the next letter, there will be a position in images where letters differ, or the difference is large enough, that image of the second morphism has to catch-up before error can be verified. Also from our formulation of $\wPCP$, it is possible that images are of equal length which means that the word is not a solution of $\wPCP$.
\setcounter{figure}{1} 
\begin{figure}
\begin{center}
\scalebox{0.8}{\input{fa2.tex}}
\end{center}
\caption{The weighted automaton $\A$.
In the figure  $a\in A$.}
\end{figure}



The following Lemma shows that for each case, there exists a path with zero weight ending in state $q_4$.

\vspace{0.5pc}
\noindent
\textbf{Lemma~\ref{SolLang}.}
\textit{Let $w\in A^\omega$. Then $w$ is a solution of an instance $(g,h)$ of the $\wPCP$ if and only if $w\notin L(\A)$.}
\begin{proof}
Assume first that $w$ is a solution to the instance $(h,g)$ of the $\wPCP$ and assume contrary to the claim that there is an accepting path of~$w$
in $\A$. 

There are three cases to be considered for the accepting path.
\begin{enumerate}[(i)]
\item An edge from $q_0$ to $q_4$ is used, or
\item the path does not visit $q_2$ or $q_3$, or
\item the path visits either $q_2$ or $q_3$.
\end{enumerate}
Assume first that $w$ is accepted by a path $\pi$ that goes to $q_4$ directly from $q_0$. To get zero weight, either $|h(w(1))|=|g(w(1))|$, meaning that $w$ is not a solution, or $j_k-c=0$ for some position $k$, but this is not possible because letters at position $k$ are equal under both morphisms. 

If the accepting path does not visit $q_2$ or $q_3$, then for some prefix $p$ $|h(p)|=|g(p)|$ which implies that $w$ is not a solution.

Finally if the path visits $q_2$, in other words $w$ has a prefix $p=auxvy$, where $x,y\in A$, such that $a$ is read using the edge $\seq{q_0,a,q_1,s(|h(u)|-|g(u)|)}$, $u$ is read in state
$q_1$ and $v$ in state $q_2$, and when reading the letter $y$ the path moves to $q_4$. The
weight $\gamma(p)$ of $p$ is now
$s(|h(au)|-|g(au)|)+s(k-|g(x)|)+j_k+s(-|g(v)|)+ (-s\ell-c)
=s\big(|h(u)|+k - |g(uxv)| -\ell\big)+j_k-c$
where $h(x)(k)=b_{j_k}$ and $g(y)(\ell)\ne b_c$. As $j_k<s$ and $c<s$, we have that
$\gamma(p)=0$ if and only if $|h(au)|+k = |g(auxv)| +\ell$ and $j_k=c$. Denote
$r=|h(au)|+k$. Now, $\gamma(p)=0$ if and only if $h(w)(r)=b_{j_k}\ne
b_c=g(w)(r)$, which is a contradiction since $w$ was assumed to be a solution of $(g,h)$. Moreover, for paths visiting $q_3$ the proof is symmetric.

For the second half of the claim, assume that $w$ is not a solution.
We summarize the possible cases for a word $w\in A^\omega$ that is not a solution of $\wPCP$. For $w$ there exists a prefix $p$ such that one of the following holds
\begin{enumerate}[(i)]
\item $|h(p)|=|g(p)|$ and $|p|=1$, or
\item $|h(p)|=|g(p)|$ and $|p|>1$, or
\item $h(p)(i)\neq g(p)(i)$ and $i$ is in the image of the first letter of $p$ under both $h$ and $g$, or
\item $h(p)(i)\neq g(p)(i)$ and $i$ is in the image of the same letter of $p$ under both $h$ and $g$, or
\item $|h(p)|>|g(p)|$ and error is in images of different letters of $p$ under $h$ and $g$, or
\item $|g(p)|>|h(p)|$ and error is in images of different letters of $p$ under $h$ and $g$.
\end{enumerate}

Assume the first case. Now $p=a$ and $w$ is accepted by using the edge $\seq{q_0,a,q_4,s(|h(a)|-|g(a)|)}=\seq{q_0,a,q_4,0}$.

Assume the second case. Now consider $pb=aub$, where $a,b\in A$ and $u\in A^*$. Using the edge $\seq{q_0,a,q_1,s(|h(a)|-|g(a)|)}$ followed by transition 
$$\seq{q_1,u(i),q_1,s(|h(u(i))|-|g(u(i))|)}$$ 
for each letter $u(i)$ of $u$ and finally transition $\seq{q_1,b,q_4,0}$ the computation reaches $q_4$. By our assumption $|h(p)|=|g(p)|$ and thus the total weight is 0.

Assume the third case. Now let the first letter of $p$ be $a$. By using the transition $\seq{q_0,a,q_4,j_k-c}$ we get an accepting computation for $w$.

Assume the fourth case. Let $p=aub$, where $a,b\in A$ and $u\in A^*$. Using the transition $\seq{q_0,a,q_1,s(|h(a)|-|g(a)|)}$ followed by transitions 
$$\seq{q_1,u(i),q_1,s(|h(u(i))|-|g(u(i))|)}$$ 
for each letter $u(i)$ of $u$ and finally transition $\seq{q_1,a,q_4,(k-\ell)s+j_k-c}$ the computation reaches $q_4$. The weight is
\begin{align*}
s(|h(au)|-|g(au)|)+s(k-\ell)+j_k-c=0
\end{align*}
when $k$ and $\ell$ are according to be the position of error in both images.

Assume the fifth case. Let $r$ be the minimal position for which $h(w)(r)\ne g(w)(r)$.
In other words for $p=c_1\cdots c_n$, there exists a position $s<n$ such that
$r=|h(c_1c_2\cdots c_{s-1})|+k$ where $k\le |h(c_{s})|$, and $r=|g(c_1c_2\cdots
c_{n-1})|+\ell$ where $\ell \le |g(c_{n})|$. Denote $h(w)(r)=b_{j_k}$. It is the
$k$th letter of the image $h(c_s)$, and $g(w)(r)$ is the $\ell$th letter of the image
$g(c_n)$. Also, these letters are nonequal.

Now, $w$ is accepted in the state $q_4$ with the following path: First $c_1$ is read with transition $\seq{q_0,c_1,q_2,s(|h(c_1)|-|g(c_1)|)}$, and the prefix
$c_2\cdots c_{t-1}$ is read in state $q_1$ with weight
$
s(|h(c_2\cdots c_{t-1})|-|g(c_2\cdots c_{t-1})|).
$
When reading $c_{t}$, the automaton uses
the error guessing transition
$
\seq{q_1,c_t,q_2,s(k-|g(c_t)|)+j_k},
$
and then the word $c_{t+1}\cdots c_{n-1}$ is read in state $q_2$ with weight
$
s(-|g(c_{t+1}\cdots c_{n-1})|).
$
Finally, while reading $c_n$, the state $q_4$ is reached by the error verifying transition
$\seq{q_2,c_n,q_4,-s\ell-j_k}$. Note that such an error
verifying transition exists as the $\ell$th letter in $g(c_n)$ is not equal to the $k$th letter
$b_{j_k}$ of $h(c_t)$. Naturally after reaching $q_4$ the
weight does not change as for all letters there are only transitions with zero weight.
Now the weight of the above path is
$s(|h(c_1\cdots c_{t-1})|-|g(c_1\cdots c_{t-1})|)+s(k-|g(c_t)|)
 +j_k+s(-|g(c_{t+1}\cdots c_{n-1})|) -s\ell-j_k
=s\big(|h(c_1\cdots c_{t-1})|+k - |g(c_1\cdots c_{n-1})| -\ell\big)
=s(r-r)=0.$
Therefore, $w$ is accepted, as claimed.

Finally the sixth case is symmetric to the fifth and is proven in the similar manner. \qed
\end{proof}

\noindent
\textbf{Theorem~\ref{UniversalFA}.}
\textit{It is undecidable whether or not $L(\A)=A^\omega$ holds for 5-state integer weighted automata $\A$ over its alphabet $A$. The automaton $\A$ is depicted in Figure~\ref{fig:FA}.}
\begin{proof}
Claim follows from Lemma~\ref{SolLang} and the undecidability of infinite $\PCP$, see \cite{R85}. The automaton $\A$ is depicted in Figure~\ref{fig:FA}. \qed
\end{proof}

\noindent
\textbf{Corollary~\ref{NonUniversalFA}.}
\textit{It is undecidable whether or not for weighted automaton $\A$, there exists a word $w\in A^\omega$ such that for its each computation path $\pi$ and prefix $p\leq\pi$, $\gamma(p)\neq0$ holds.}
\begin{proof}
The statement formulates the condition for non-universality. By previous Theorem, universality problem is undecidable, and thus so is its complement problem. \qed
\end{proof}

More precisely, we have an infinite subclass of integer weighted automata corresponding to $\wPCP$. We do not know whether there exists weighted automaton not in this subclass for which universality problem is undecidable.

\vspace{0.5pc}
\noindent
\textbf{Theorem~\ref{UndecidableReverseAcceptance}.}
\textit{It is undecidable whether or not $L_R(\B)=A^\omega$ holds for 5-state integer weighted automata $\B$ over its alphabet $A$.}
\begin{proof}
Consider automaton $\A=(Q,A,\sigma,q_0,\{q_4\})$ of previous Theorem. We construct automaton $\B=(Q,A,\sigma',q_4,\{q_0\})$ by reversing all transitions, i.e. if $\seq{q,a,p,z}\in\sigma$ then $\seq{p,a,q,-z}\in\sigma'$.

Now if there is a prefix $p$ of computation path $\pi$ of word $w\in A^\omega$ of $\A$. Let 
\begin{align*}
p&=\seq{q_0,a_1,q_1,z_1}\seq{q_1,a_2,q_1,z_2}\cdots\seq{q_1,a_k,q_1,z_k}\seq{q_1,a_{k+1},q_2,z_{k+1}} \\
 &\phantom{=} \seq{q_2,a_{k+2},q_2,z_{k+2}}\cdots\seq{q_2,a_\ell,q_4,z_\ell}\seq{q_4,a_{\ell+1},q_4,0}\cdots\seq{q_4,a_n,q_4,0}
\end{align*}
and $\gamma(p)=z_1+\ldots+z_n=z$.

Now in $\B$, there is a computation path for $p^R$ of computation path $\pi$ of word $w\in A^\omega$ of $\B$.
\begin{align*}
p^R&=\seq{q_4,a_n,q_4,0}\cdots\seq{q_4,a_{\ell+1},q_4,0}\seq{q_4,a_\ell,q_2,-z_\ell}\cdots\seq{q_2,a_{k+2},q_2,-z_{k+2}} \\
 &\phantom{=} \seq{q_2,a_{k+1},q_1,-z_{k+1}}\seq{q_1,a_k,q_1,-z_k}\cdots\seq{q_1,a_2,q_1,-z_2}\seq{q_1,a_1,q_0,-z_1}
\end{align*}
with weight $\gamma(p^R)=-z_n-\ldots-z_1=-z$.

In similar way we can show that other computation paths in $\A$ have corresponding path in $\B$.

Clearly $w$ is accepted by $\A$ with $\gamma(p)=0$ if and also if $w$ is accepted by $\B$ with $\gamma(p^R)=0$. \qed
\end{proof}

Note that the automaton of Theorem~\ref{UniversalFA} generates infinite words while the automaton of Theorem~\ref{UndecidableReverseAcceptance} is a finite automaton that verifies that proposed infinite word is accepted into the language.

\begin{figure}
\begin{center}
\scalebox{0.85}{\input{fa_reversed.tex}}
\end{center}
\caption{The weighted automaton $\B$.
In the figure  $a\in A$.}
\end{figure}

\newpage
\section*{Appendix B}
\textbf{Theorem~\ref{weighted Word Game}.}
\textit{It is undecidable whether Attacker has a winning strategy in the weighted Word Game with words over a binary free group alphabet.}
\begin{proof} 
The proof is based on the reduction of the the universality problem for
weighted automata on infinite words to the problem of checking a winning strategy in the weighted Word Game. 

Let us modify the automaton $\A$ without changing the language which it accepts. We will remove 
local cycles from states $q_1,q_2,q_3$ and $q_4$, then we will make a copy of such automaton
having states $q'_1,q'_2,q'_3$ and $q'_4$ corresponding to $q_1,q_2,q_3$ and $q_4$ and adding 
edges from $q_i$ to $q'_i$ and $q'_i$ to $q_i$ with a label $(a,x)$ if the original automaton $\A$ had a cycle from $q_i$ to  $q_i$
with a label $(a,x)$. Finally we rename states  $q'_1,q'_2,q'_3$ and $q'_4$ into $q_5,q_6,q_7$ and $q_8$ for convenience.
Let us define the following initial instance of the weighted Word Game: 
\begin{itemize} 
\item Defender's words are just single letters from the alphabet $A$ with weight 0 and 
\item there are three types of words of Attacker. Either the word $\#$ with weight 0, or $\overline{a}\cdot \overline{q_i}$ with weight $x$ that encodes a single transition of our modified $\A$ from a state $q_0$ to $q_i$ with a symbol $a \in A$ and a weight $x \in \Z$, or $\overline{a} \cdot q_i  \cdot \overline{\#} \cdot  \overline{q_j}$ with weight $x$ that encodes a single transition from $q_i\neq q_0$ to $q_j$ with a symbol $a\in A$ and weight $x\in\Z$. Attacker's words are over a group alphabet $\Gamma=A \cup A^{-1} \cup Q \cup Q^{-1}\cup \{\#,\overline{\#}\}$.
\end{itemize} 

The initial word is $q_0$. Then in the above game Defender can avoid reaching a configuration $q_0\overline{q_4}$ or $q_0\overline{q_8}$ with weight 0 if and only if 
there is an infinite word that is not accepted by the weighted 
automaton $\A$.
Let us assume that there is an infinite word $w$ that is not accepted by $\A$.
Defender can make a choice about every next symbol and can generate the word $w$ during the play.
At some point Attacker may choose to check whether played prefix $p$ of $w$ can be accepted by $\A$ and starts canceling letters of $p$ with words that have the correct prefix and also follows correctly the transitions according to the state structure of automaton  $\A$.

In this case any alternating play of Defender and Attacker 
with a sequence of visited states $q_{0},q_{i_1},\ldots ,q_{i_t}$ 
in the correct order by reading a word $a_{i_1}, a_{i_2},\ldots , a_{i_{t}}$  with the corresponding weights 
$x_{i_1},  x_{i_2},\ldots , x_{i_{t}}$ give us the following reduced word
\begin{align*}
q_0\cdot a_{i_1}\cdot\#\cdot\ldots\cdot a_{i_t}\cdot \underbracket{\overline{a_{i_t}}\cdot\overline{q_{i_1}}}\cdot a_{j_1}\cdot \underbracket{\overline{a_{j_1}}\cdot q_{i_1}\cdot \overline{\#}\cdot\overline{a_{i_{t-1}}}\cdot \overline{q_{i_2}}}\cdot\ldots\cdot \\
a_{j_t}\cdot \underbracket{\overline{a_{j_t}}\cdot q_{i_t}\cdot \overline{\#}\cdot\overline{a_{i_{1}}}\cdot \overline{q_{i_t}}} = q_0\overline{q_{i_t}}
\end{align*}
and the weight $0+0+\ldots+0+ x_{i_t}+\ldots+ x_{i_1}$
which will always be a nonzero if Defender follows the word $w$ that is not accepted by the 
automaton $\A$. Note that letters $a_{j_k}$, played by Defender after Attacker has started simulating the automaton, play no role in the computation.

If a symbol (of Defender) $a$ is followed by a word ( of Attacker) with a prefix 
$\overline{b}$,  where $a \neq b$ then the configuration of this game will contain non-cancelling factor $a\overline{b}$
due to the fact the all symbols in the encodings are from the free group.
In the same way if the symbol of Defender was correctly followed by Attacker but the order of applied transitions does not
correspond to the state structure of automaton  $\A$ the 
configuration of a game will contain non-cancelling factor $\overline{q_i}q_j$ for some $i,j$ such that $i \neq j$.
Therefore in both cases the configurations $q_0\overline{q_4}$ and $q_0\overline{q_8}$ are not reachable by Attacker.  

In the opposite direction we assume that $q_0\overline{q_4}$ (or $q_0\overline{q_8}$) is reachable by Attacker
then the middle part of the concatenated words  between $q_0$ and $\overline{q_4}$ (or $q_0$ and $\overline{q_8}$)
should be equal to identity and corresponds to correct cancellation of state symbols from $Q \cup Q^{-1}$ 
that in its turn forming a path with full cancellation of  weights corresponding to weights in the original automaton.

In order to get a game where the word of a winning configuration is an empty word rather than  $q_0\overline{q_4}$ or $q_0\overline{q_8}$ we need to 
have an extra moves for Attacker and to make sure that no false solutions are added. The simple construction
of adding words $\overline{a} \cdot q_4    \cdot  \overline{q_0}$ and $\overline{a} \cdot q_8    \cdot  \overline{q_0}$ for all $a \in A$ creates no new solutions as there is no way to reach $\varepsilon,$ after $q_0$ has been canceled out.

In order to complete the proof, we will require the encoding of Lemma~\ref{groupEnc} between words over an arbitrary group alphabet and a binary group alphabet,
which is well known from the literature \cite{BM07,Paul_Igor_FI_2012}.
The Lemma's morphism gives a way to map words from an arbitrary sized group alphabet into the set of
 words over a free group alphabet with only two symbols. \qed

\end{proof}

We now modify the game of previous Theorem by encoding counter as a separate word over unary group alphabet $\Gamma'=\{\rho,\overline{\rho}\}$.

Now \emph{Word Game}  consists of Attacker and Defender having sets of words  $\{(u_1,u'_1),\ldots,(u_r,u'_r)\}\subseteq \Gamma^*\times\Gamma'^*$ and $\{(v_1,\varepsilon),\ldots,(v_s,\varepsilon)\}\subseteq\Gamma^*\times\Gamma'^*$, where $\Gamma$ is binary group alphabet.
Now the configuration of a game after $t$ rounds is a word  $w_t=(v_{i_1},\varepsilon) \cdot (u_{i_1},u'_{i_1}) \cdot (v_{i_2},\varepsilon) \cdot (u_{i_2},u'_{i_2}) \cdot \ldots \cdot (v_{i_t},\varepsilon) \cdot (u_{i_t},u'_{i_t})$, where $(u_{i_j},u'_{i_j})$ and $(v_{i_j},\varepsilon)$ are words from defined above sets of Attacker and Defender.  The initial position is the word $(w,\varepsilon)$ and target position of this game is an empty word $(\varepsilon,\varepsilon)$.
The decision problem for the word game is to check whether there exists a winning
strategy for Attacker to reach an empty word $(\varepsilon,\varepsilon)$.

\vspace{0.5pc}
\noindent
\textbf{Theorem~\ref{Word Game}.}
\textit{It is undecidable whether Attacker has a winning strategy in the Word Game with one component of words over a binary free group alphabet and the other over unary group alphabet.}
\begin{proof} 
The proof is based on Theorem~\ref{weighted Word Game} with encoding of counter as a word over unary group alphabet. 

Let us define the following initial instance of the Word Game: 
\begin{itemize} 
\item Defender's words are just single letters from the alphabet $A\times\{\varepsilon\}$ and 
\item there are three types of words of Attacker. Either the word $(\#,\varepsilon)$, or $(\overline{a}\cdot \overline{q_i},\rho^x)$ that encodes a single transition of our modified $\A$ from a state $q_0$ to $q_i$ with a symbol $a \in A$ and a weight $x \in \Z$, or $(\overline{a} \cdot q_i  \cdot \overline{\#} \cdot  \overline{q_j},\rho^x)$ that encodes a single transition from $q_i\neq q_0$ to $q_j$ with a symbol $a\in A$ and weight $x\in\Z$. Attacker's words are over a group alphabet $\Gamma\times\Gamma'$=$(A \cup A^{-1} \cup Q \cup Q^{-1}\cup \{\#,\overline{\#}\}) \times \{\rho, \overline{\rho}\}$.
\end{itemize} 

The initial pair of words is $(q_0,\varepsilon)$. Applying morphism of Lemma~\ref{groupEnc}, we have words over binary group alphabet in the first component. It is clear that the winner in weighted Word Game is also the winner in Word Game. \qed
\end{proof}

We follow idea of \cite{BP10} to construct a word game where the initial word is $(\varepsilon,\varepsilon)$. For this we do 4 consequent games over 4 disjoint group alphabets. The games are constructed in such way that $(\varepsilon,\varepsilon)$ is reached if and only if $(\varepsilon,\varepsilon)$ is reached in every game. If none of the games result in $(\varepsilon,\varepsilon)$, then there are at least 4 non-canceled words. Now if the computation is completed twice (i.e. 8 games have been played in total), the number of non-canceled elements cannot decrease.

We extend the domain of the game and a set of rules to the class of linear transformations on integer lattice $\Z^4$ and 
to the domain of braids considering moves of the game as a concatenation of braids in $B_3$ (a class of braids with only three strands) and $B_5$ (a class of braids with only five strands)
\cite{P-FSTTCS13}.

A \emph{Matrix Game} consists of two players, \emph{Attacker} and \emph{Defender} having sets of
linear transformations  $MU_1,\ldots,MU_r\subseteq \Z^{n \times n }$ and $MV_1,\ldots,MV_s\subseteq\Z^{n \times n }$
respectively and an \emph{initial vector} $\bx_0 \in \Z^n$ of the game representing a starting position. 
The \emph{dimension} of the game is clearly the dimension of the integer lattice $n$
. 
Starting from $\bx_0$, players move the current point by applying available linear transformations (by matrix multiplication)
from their respective sets in turns. 
%
The decision problem of the {\sl Matrix game} is to check whether there exist a winning
strategy for Attacker to return to the starting point (vector in $\Z^n$) of the game. 

\vspace{0.5pc}
\noindent
\textbf{Theorem~\ref{matrixgame}.}
\textit{Given two finite sets matrices $MU_1,MU_2,\ldots,MU_r\subseteq \Z^{n \times n }$ and $MV_1,MV_2,\ldots,MV_s\subseteq\Z^{n \times n }$
for \emph{Attacker} and \emph{Defender} players respectively and initial starting vector $\bx_0 \in \Z^n$.
It is undecidable whether Attacker has a winning strategy in the Matrix Game of dimension four, i.e. when $n=4$.}
\begin{proof}

Let $\Sigma_2=\{c,d,\overline{c},\overline{d}\}$ be a binary group alphabet and define $f: \Sigma_2^* \to \mathbb{Z}^{2 \times 2}$ by:
$
f(c) = \begin{psmallmatrix}1 & 2 \\ 0 & 1 \end{psmallmatrix},
f(\overline{c}) = \begin{psmallmatrix} 1 & -2 \\ 0 & 1 \end{psmallmatrix},
f(d) = \begin{psmallmatrix} 1 & 0 \\ 2 & 1 \end{psmallmatrix},
f(\overline{d}) = \begin{psmallmatrix} 1 & 0 \\ -2 & 1 \end{psmallmatrix}.
$

Then mapping $f$ is a monomorphism \cite{Paul_Igor_FI_2012} and $f(\varepsilon)$ corresponds to the identity matrix in $\Z^{2\times2}$ .
Let $\alpha$ be a function defined in Lemma~\ref{groupEnc} then by the following straightforward matrix multiplication
we have:
$$f(\alpha(z_j)) = f(c^j d\overline{c}^{j}) = 
\begin{pmatrix}1+4j & -8j^2 \\ 2 & 1-4j\end{pmatrix}.
$$
Note that when we multiply several encoded letters, in the bottom right corner we always have $1\pmod 4$.

Let us show that if $\begin{psmallmatrix}
1 & 0\end{psmallmatrix}M=\begin{psmallmatrix}
1 & 0\end{psmallmatrix}$, where $M$ is an image of a word over binary group alphabet under $f$, that is $M\in\{f(\alpha(w))\mid w\in\Gamma^*\}$. Then $M$ is the identity matrix. The reasoning follows \cite{Paul_Igor_FI_2012}. Let $M=\begin{psmallmatrix}
m_{11} & m_{12} \\ m_{21} & m_{22}\end{psmallmatrix}$, now $\begin{psmallmatrix}
1 & 0\end{psmallmatrix}M=\begin{psmallmatrix}
m_{11} & m_{12}\end{psmallmatrix}$ which implies that $m_{11}=1$ and $m_{12}=0$. By the previous observation $m_{22}=1$. The final letter of $\alpha(w)$ is $\overline{c}$ which is $\begin{psmallmatrix}1 & -2 \\ 0 & 1\end{psmallmatrix}$ under $f$. Let $Y=f(\alpha(w))f(\overline{c})^{-1}=\begin{psmallmatrix}x & y \\ z & v\end{psmallmatrix}$. Now $f(\alpha(w))=\begin{psmallmatrix}x & y \\ z & v\end{psmallmatrix}\begin{psmallmatrix}1 & -2 \\ 0 & 1\end{psmallmatrix}=\begin{psmallmatrix}x & y-2x \\ z & v-2z\end{psmallmatrix}$. Since $x=1$, $y-2x=0$ and $x-2x=1$, we see that $f(\alpha(w))=\begin{psmallmatrix}1 & 0 \\ z & 1\end{psmallmatrix}=f(d)^{z/2}$ but by definition of encoding this is possible only when $z=0$. This implies that $f(\alpha(w))$ is the identity matrix.

Let us encode the {\sl Word Game} into the {\sl Matrix Game}. We construct $4\times4$ matrices with words of the first component encoded by $f$ in the upper left corner and words from the second components encoded by $f$ in the lower right corner. The direct application of the above function to the group words
of the {\sl Word Game} gives us a set of matrices for Attacker and a set of matrices for Defender from $SL(4,\Z)$ (i.e. integer matrices of dimension $4$ with the determinant $1$), where Attacker can only win if the following moves of the game 
for every infinite run at some point the identity matrix is reachable after a move of Attacker.
By previous considerations for vector $x_0=(1,0,1,0)$, the following equation $x_0=M \cdot x_0$, where $M\in SL(4,\Z)$
has only one matrix  $M$ satisfying the above statement, the identity matrix in $\Z^4$. Therefore for every
Matrix Game where the initial vector $x_0$ is $(1,0,1,0)$ the question about the winning 
strategy of reaching $x_0$ is equivalent to the question of reaching the identity matrix in the product with 
alternation in applications of Defender's and Attacker's linear transformations, which in its turn corresponds to reaching an empty word 
in the {\sl Word Game}. \qed
\end{proof}

Now we translate the Attacker-Defender games into games on topological objects  - braids in $B_n$.  
Braids are classical topological objects that attracted a lot of attention
due to their connections to topological knots and links 
as well as their applications to polymer chemistry, molecular biology,
cryptography, quantum computations and robotics \cite{GC2006,Dehornoy, EPCHLT92,G,PP2011}.
There is also recent interest about the complexity and termination of the games on braids \cite{CDW, BC-preprint}
that are defined with specific rules of adding and removing crossing. In this paper we consider very simple games on braids with only 3 or 5 strands (i.e. $B_3$ or $B_5$) where the braid is modified by composition with a finite set of braids.  We show that it is undecidable to check the existence of a winning strategy in such game, while the reachability with a single player (i.e. with nondeterministic concatenation from a single set) was shown to be decidable for $B_3$ and undecidable for $B_5$ in \cite{P-FSTTCS13}.

\noindent
\textbf{Definition~\ref{Braid_Artin}.}
\textit{The $n$-strand braid group  $B_n$ is the group given by the presentation with $n-1$
generators $\sigma_1 , \ldots , \sigma_{n-1}$ and the following relations
$\sigma_i \sigma_j = \sigma_j \sigma_i$, for $|i-j| \geq 2$
and $\sigma_i \sigma_{i+1} \sigma_i = \sigma_{i+1} \sigma_i \sigma_{i+1}$
for $1 \leq i \leq n-2$. These relations are called Artin's relations.
%
Words in the alphabet $\{ {\sigma}$ , ${\sigma}^{-1} \}$ will be referred to as braid words}
\footnote{ Whenever a crossing of strands $i$ and $i + 1$ is encountered, $\sigma_i$ or ${\sigma_i}^{-1}$  
is written down, depending on whether strand $i$ moves under or over strand $i + 1$.}.

The \emph{fundamental braid} of $B_n$ is 
$$\Delta_n=(\sigma_{n-1}\sigma_{n-2}\ldots\sigma_1)(\sigma_{n-1}\sigma_{n-2}\ldots\sigma_2)\ldots\sigma_{n-1}.$$
Geometrically, the fundamental braid is obtained by lifting the bottom ends of the identity
braid and flipping (right side over left) while keeping the ends of the strings in a line.

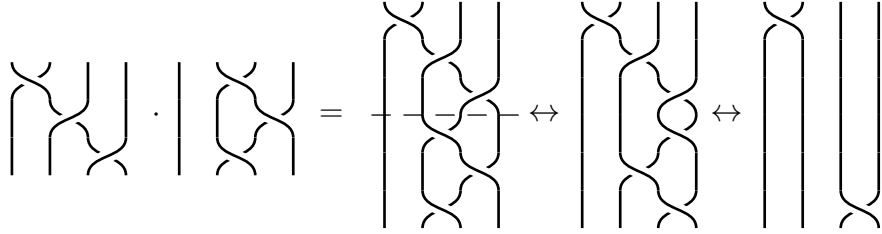
\begin{center}
\begin{tikzpicture}
\braid[strands=4,braid start={(-1.2,-0.8)}]
{\sigma_1 \sigma_2^{-1} \sigma_3^{-1}}
\node[font=\large] at (1.2,-1.5) {\(\cdot \)};
\braid[strands=4,braid start={(1,-0.8)}]
{\sigma_2 \sigma_3 \sigma_2}
\node[font=\large] at (3.5,-1.5) {\(= \)};
\braid[strands=4,braid start={(3.7,0)}]
{\sigma_1 \sigma_2^{-1} \sigma_3^{-1} \sigma_2 \sigma_3 \sigma_2 }
\node[font=\large] at (5,-1.5) {\( - - - - - \)};
\node[font=\large] at (6.3,-1.5) {\(\leftrightarrow \)};
\braid[strands=4,braid start={(6.3,0)}]
{\sigma_1 \sigma_2^{-1} \sigma_3^{-1} \sigma_3 \sigma_2 \sigma_3 }
\node[font=\large] at (8.7,-1.5) {\(\leftrightarrow \)};
\braid[strands=4,braid start={(8.7,0)}]
{\sigma_1  \sigma_5  \sigma_5 \sigma_5 \sigma_5 \sigma_3 }
\end{tikzpicture}
\end{center}

The Braid Game can be defined in a way, where a set of  words for Attacker $\{a_1,...,a_r\} \subseteq B_n$ and Defender
$\{d_1,...,d_s\} \subseteq B_n$ will correspond to braids (or braid words in $B_n$). The game is starting with a braid and each
following configuration of the game is changed by Attacker or Defender by concatenating braids from their corresponding sets.
Given two geometric braids, we can compose them, i.e. put one after the other making the endpoints of the first one coincide with the starting points of the second one. There is a neutral element for the composition: it is the trivial braid, also called identity braid, i.e. the class of the geometric braid where all the strings are straight. 
Two geometric braids are isotopic if there is a continuous deformation of the ambient space that
deforms one into the other, by a deformation that keeps every point in the two bordering planes fixed.

Finally, the goal of Attacker is to unbraid, i.e. to reach a configuration of the game that is isotopic to the trivial braid (empty word) and 
Defender tries to keep Attacker from reaching it.
Two braids are isotopic if their braid words can be translated one into each other via the relations from the Definition~\ref{Braid_Artin}
plus the relations $\sigma_i \sigma_i^{-1}  = \sigma_i^{-1} \sigma_i =1$, where $1$ is the identity (trivial braid).

\vspace{0.5pc}
\noindent
\textbf{Theorem~\ref{braidgame3}.}
\textit{The Braid Game is undecidable for braids from $B_3$ starting from non-trivial braid and for braids from $B_5$ starting from a trivial braid.}
\begin{proof}

Let $\Sigma_2=\{c,d,\overline{c},\overline{d}\}$ be a binary group alphabet and define $f: \Sigma_2^* \to {B}_3$ by:
$
f(c) = {{\sigma}_1}^4,
f(\overline{c}) = {{\sigma}_1}^{-4},
f(d) = {{\sigma}_2}^4,
f(\overline{d}) = {{\sigma}_2}^{-4}.
$
Then mapping $f$ is a monomorphism \cite{BD99}.
The above two morphisms give a way to map words from an arbitrary sized alphabet into the set of
braid words in $B_3$.

Let $\alpha$ be the mapping from Lemma~\ref{groupEnc} then:
$$f(\alpha(z_j)) = f(c^j d\overline{c}^{j}) = {{\sigma}_1}^{4j} {{\sigma}_2}^4 {{\sigma}_1}^{-4j}
$$
and the length of a braid word from $B_3$ corresponding to a symbol $z_j \in  \Sigma'$  is $8j+4$.
Now we again can use the {\sl weighted Word Game} as any word over a binary group alphabet can be uniquely mapped into a braid,
where empty word will correspond to a braid which is isotopic to the trivial braid and concatenation of words over group alphabet 
corresponds to concatenation of braids in $B_3$. The counter $x\in\Z$ is mapped into a braid word $\Delta_3^{2x}$, where $\Delta_3^2=(\sigma_1\sigma_2\sigma_1)^2$ is a central element of $B_3$. 

Subgroups $\langle \sigma_1^4,\sigma_2^4\rangle, \langle\sigma_4^2,d\rangle$ of the group $B_5$ are free and $B_5$ contains the direct product $\langle \sigma_1^4,\sigma_2^4\rangle\times \langle\sigma_4^2,d\rangle$ of two free groups of rank 2 as a subgroup, where $d = \sigma_4 \sigma_3 \sigma_2 \sigma_1^2 \sigma_2 \sigma_3 \sigma_4$ \cite{BD99}. Now we can uniquely encode pair of words of {\sl Word Game} into $B_5$. Using the Word Game, where the initial word is $(\varepsilon,\varepsilon)$, we can construct a Braid Game from $B_5$ starting from trivial braid. \qed
\end{proof}

\newpage
\section*{Appendix C}
Robot Games is another simple Attacker-Defender game \cite{RobotGames} in which deciding whether Attacker has a winning strategy is believed to be undecidable for $n=9$. In Robot Game, players are given vector sets $U=\{\bu_1,\ldots,\bu_k\},V=\{\bv_1,\ldots,\bv_\ell\}$ for Attacker and Defender, respectively, initial vector $\bx$ and target vector $\mathbf{y}$. Each turn players add a vector from their respective vector sets to current configuration with Attacker trying to reach $\mathbf{y}$ and Defender keep Attacker from reaching it.

By encoding $n$-dimensional Robot Game into matrices, we get Matrix Game of dimension $2n$ for which it is undecidable whether Attacker has a winning strategy. Our proof gives significantly smaller dimension.

\begin{corollary}
If deciding whether Attacker has a winning strategy in Robot Game of dimension $n$, then it is undecidable whether Attacker has a winning strategy in the Matrix Game of dimension $2n$.
\end{corollary}
\begin{proof}
Let $U=\{\bu_1,\ldots,\bu_k\}$ and $V=\{\bv_1,\ldots,\bv_\ell\}$ be Attacker's and Defender's vector sets, initial vector $\bx=(x_1,\ldots,x_n)$, target vector $\mathbf{y}=(y_1,\ldots,y_n)$ in Robot Games of dimension $n$. Let $d$ be a mapping $\Z^n\to\Z^{n\times n}$ that puts components of a vector on the main diagonal of $n\times n$ matrix. In Matrix Game for each vector $\bv_i$, Defender has $2n\times 2n$ matrix $V_i$ with $d(1,\ldots,1)$ in top left and bottom right corners and $d(\bv_i)$ in top right corner, and Attacker has similarly defined $U_j$ matrices. Let $(x_1,\ldots,x_n,1,\ldots,1)^T\in\Z^{2n}$ be a vector corresponding to initial vector of Robot Game. Now vector $(y_1,\ldots,y_n,1,\ldots,1)^T\in\Z^{2n}$ can be reached in the Matrix Game if and only if $(y_1,\ldots,y_n)\in\Z^n$ can be reached in Robot Game. \qed
\end{proof}

\end{document}